\documentclass[11pt]{cernrep}
\usepackage{hyperref,cite}
\usepackage[latin1]{inputenc}
\usepackage[dvips]{graphicx,epsfig,color}
\usepackage{wrapfig,rotating}
\usepackage{amssymb,amsmath}

\newcommand{\newc}{\newcommand}
\newc{\hwpp}{\textrm{Herwig++}}
\newc{\ue}{underlying event}
\newc{\GeV}{\, {\rm GeV}}
\newc{\pt}{p_t}
\newc{\ptmin}{\pt^{\rm min}}

\begin{document}
\title{Underlying events in \hwpp \footnote{{}\hspace{0.5em}to appear in the
    proceedings of the HERA and the LHC workshop.}
\hfill\parbox[b]{4cm}{\rm \footnotesize\raggedleft
CERN-PH-TH-2008-195\\KA-TP-22-2008\\MCnet/08/11}}
\author{
  Manuel~B\"ahr$^{1}$, Stefan~Gieseke$^{1}$ and
  Michael~H.~Seymour$^{2}$
}
\institute{
  $^{1}$\ Institut f\"ur Theoretische Physik, Universit\"at Karlsruhe,\\
  $^{2}$\ Physics Department, CERN, and\\
  \hspace{0.14cm} School of Physics and Astronomy, University of Manchester.
}
\maketitle
\begin{abstract}
  In this contribution we describe the new model of multiple partonic
  interactions (MPI) that has been implemented in \hwpp. Tuning its two
  free parameters is enough to find a good description of CDF underlying
  event data. We show extrapolations to the LHC and compare them to
  results from other models.
\end{abstract}

\section{Introduction}

With the advent of the Large Hadron Collider (LHC) in the near future it
will become increasingly important to gain a detailed understanding of
all sources of hadronic activity in a high energy scattering event.  An
important source of additional soft jets will be the presence of the
underlying event. From the experimental point of view, the underlying
event contains all activity in a hadronic collision that is not related
to the signal particles from the hard process, e.g.\ leptons or missing
transverse energy. The additional particles may result from the initial
state radiation of additional gluons or from additional hard (or soft)
scatters that occur during the same hadron--hadron collision. Jet
measurements are particularly sensitive to the underlying event because,
although a jet's energy is dominated by the primary hard parton that
initiated it, jet algorithms inevitably gather together all other energy
deposits in its vicinity, giving an important correction to its energy
and internal structure.

In this note, based on Ref.~\cite{Bahr:2008dy}, we want to focus on the
description of the hard component of the underlying event, which stems
from additional hard scatters within the same proton. Not only does this
model give us a simple unitarization of the hard cross section, it also
allows to give a good description of the additional substructure of the
underlying events. It turns out that most activity in the underlying
event can be understood in terms of hard minijets. We therefore adopt
this model, based on the model \textsf{JIMMY} \cite{Butterworth:1996zw},
for our new event generator \hwpp~\cite{Bahr:2008pv}.

An extension to this model along the lines of \cite{Borozan:2002fk},
which also includes soft scatters is underway and will most probably be
available for the next release of \hwpp. Covering the entire $\pt$ range
will also allow us to describe minimum bias interactions.  We have
examined the parameter space of such models at Tevatron and LHC energies
in Ref.~\cite{Bahr:2008wk}. Existing measurements and the possible range
of LHC measurements are used there to identify the maximally allowed
parameter space.

\section{Tevatron results}

We have performed a tune of the model by calculating the total $\chi^2$
against the jet data ($\pt^{\rm ljet} > 20 \GeV$) from
Ref.~\cite{Affolder:2001xt}. For this analysis each event is partitioned
into three parts, the \textbf{towards, away} and \textbf{transverse}
regions. These regions are equal in size in $\eta - \phi$ space and
classify where particles are located in this space with respect to the
hardest jet in the event. We compare our predictions to data for the
average number of charged particles and for the scalar $\pt$ sum in each
of these regions.

\begin{figure}[t]
  \begin{center}
    \includegraphics[width=0.45\columnwidth]{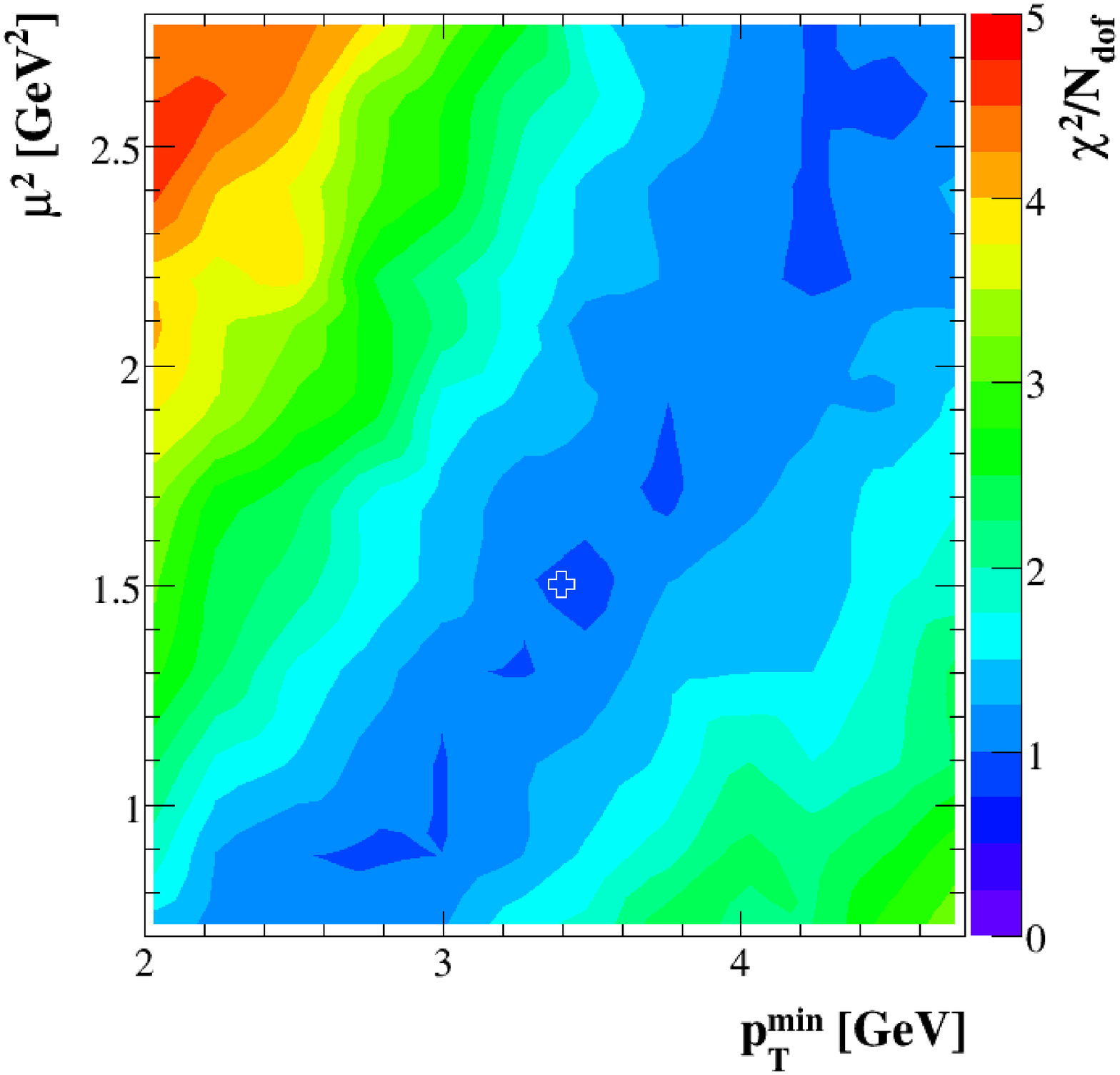}
    \includegraphics[width=0.45\columnwidth]{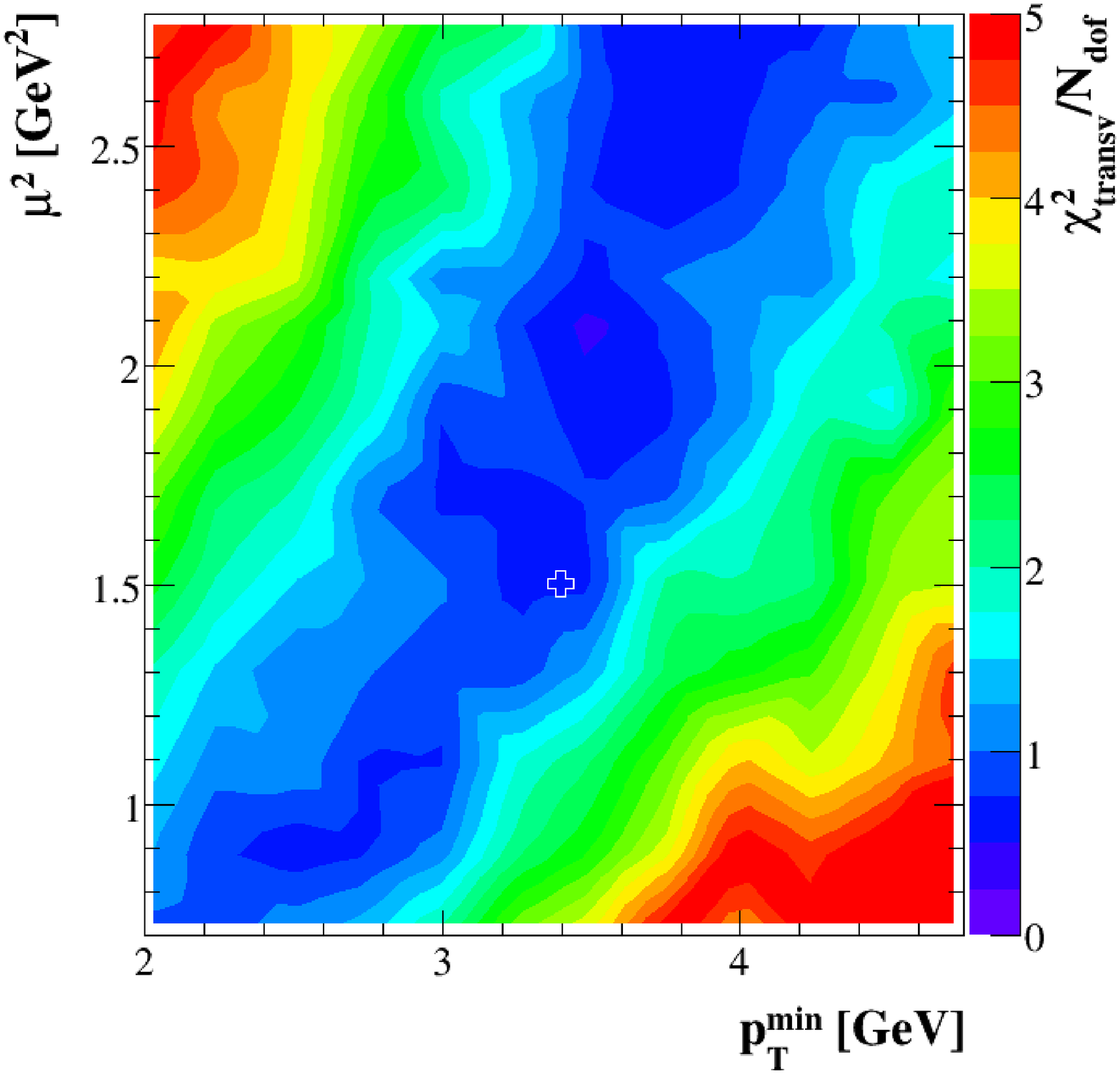}
    \label{Fig:map}
    \caption{ Contour plots for the $\chi^2$ per degree of freedom of all
      discussed observables (left) and only the ones from the transverse
      region (right). The cross indicates the location of our preferred
      tune.}
  \end{center}
\end{figure}

The parameter space for this tune is two dimensional and consists of the
$\pt$ cutoff $\ptmin$ and the inverse hadron radius squared, $\mu^2$. In
Fig.~\ref{Fig:map} we show the $\chi^2$ contour for describing all six
observables and especially those from the transverse region, which is
particularly sensitive to the underlying event. For these, and all
subsequent plots, we have used \hwpp\ version 2.2.1 and the built-in
MRST~2001 LO\cite{Martin:2001es} PDFs. All parameters, apart from the
ones we were tuning, were left at their default values.

The description of the Tevatron data is truly satisfactory for the
entire range of considered values of $\ptmin$. For each point on the
$x$-axis we can find a point on the $y$-axis to give a reasonable
fit. Nevertheless an optimum can be found between 3 \ldots\ 4 GeV. The
strong and constant correlation between $\ptmin$ and $\mu^2$ is due to
the fact that a smaller hadron radius will always balance against a
larger $\pt$ cutoff as far as the underlying event activity is
concerned. As a default tune we use $\ptmin = 3.4 \GeV$ and $\mu^2 = 1.5
\GeV^2$, which results in an overall $\chi^2/N_{\rm dof}$ of 1.3.

\section{LHC extrapolation}

\begin{wrapfigure}{r}{0.5\columnwidth}
  \vspace*{-0.5cm}
  \centerline{
    \includegraphics[width=.43\columnwidth]{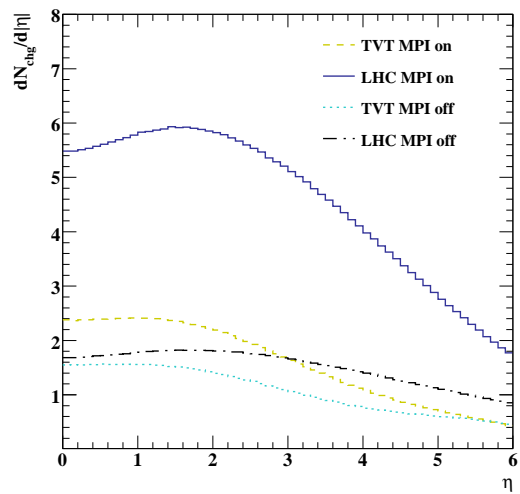}  }
  \vspace*{-0.4cm}
  \caption{
    \label{fig:obs_lhc}
    Differential multiplicity distribution with respect to $|\eta|$. The
    different data sets are: Tevatron with MPI off, LHC with MPI off,
    Tevatron with MPI on and LHC with MPI on.}
\end{wrapfigure}

We start the discussion of our predictions for the LHC with the plot in
Fig.~\ref{fig:obs_lhc}. The plot shows the mean charged multiplicity as
a function of pseudorapidity, $\eta$. We show \hwpp\ with and without
MPI. We used QCD jet production with a minimal $p_T$ of 20~GeV as signal
process. The MPI parameters were left at their default values, i.e.\ the
fit to Tevatron CDF data. The effect of MPI is clearly visible, growing
significantly from the Tevatron to the LHC.

\begin{figure}[t]
  \includegraphics[%
    width=.48\textwidth,keepaspectratio]{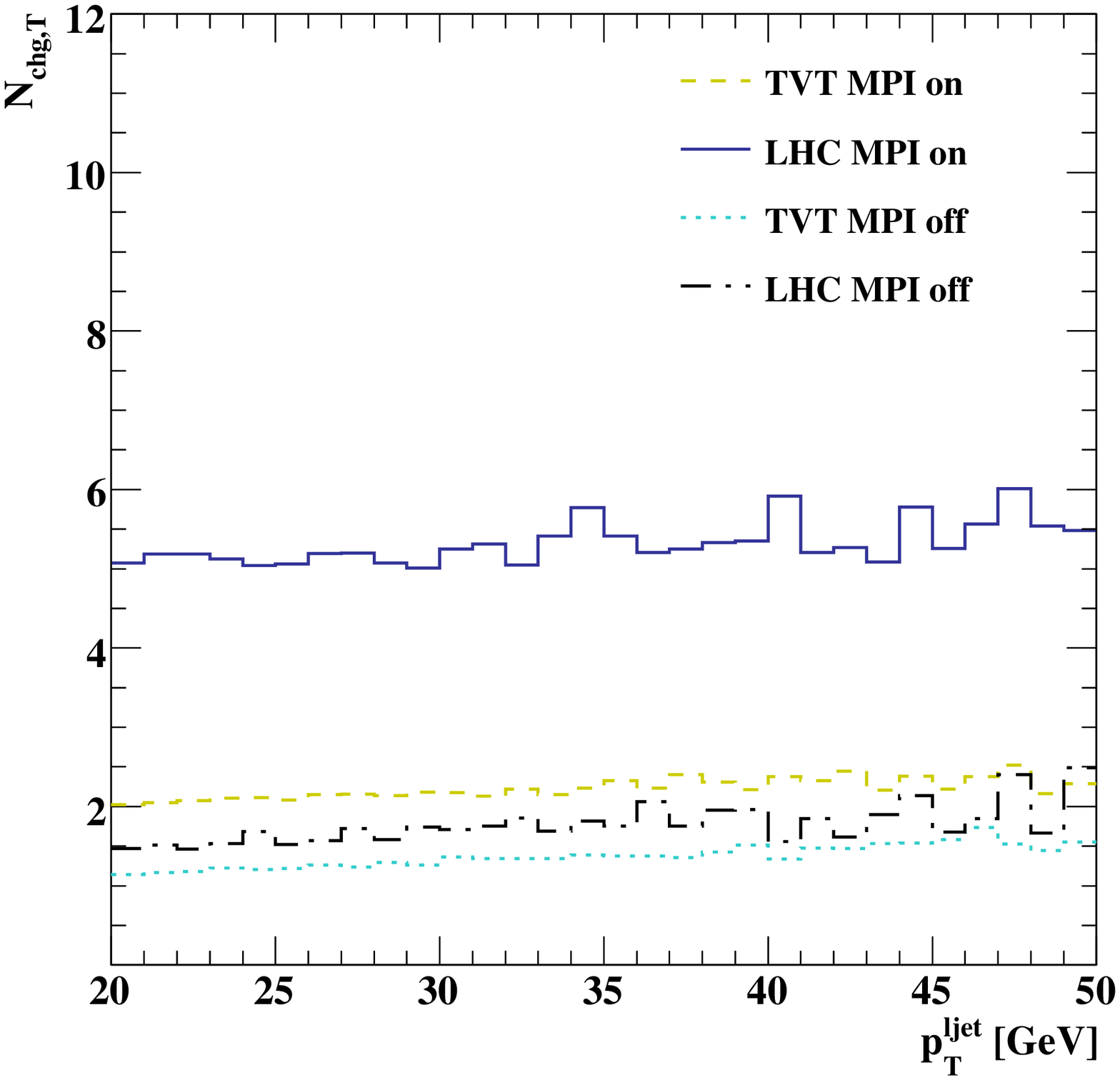}
  \includegraphics[%
    width=.48\textwidth,keepaspectratio]{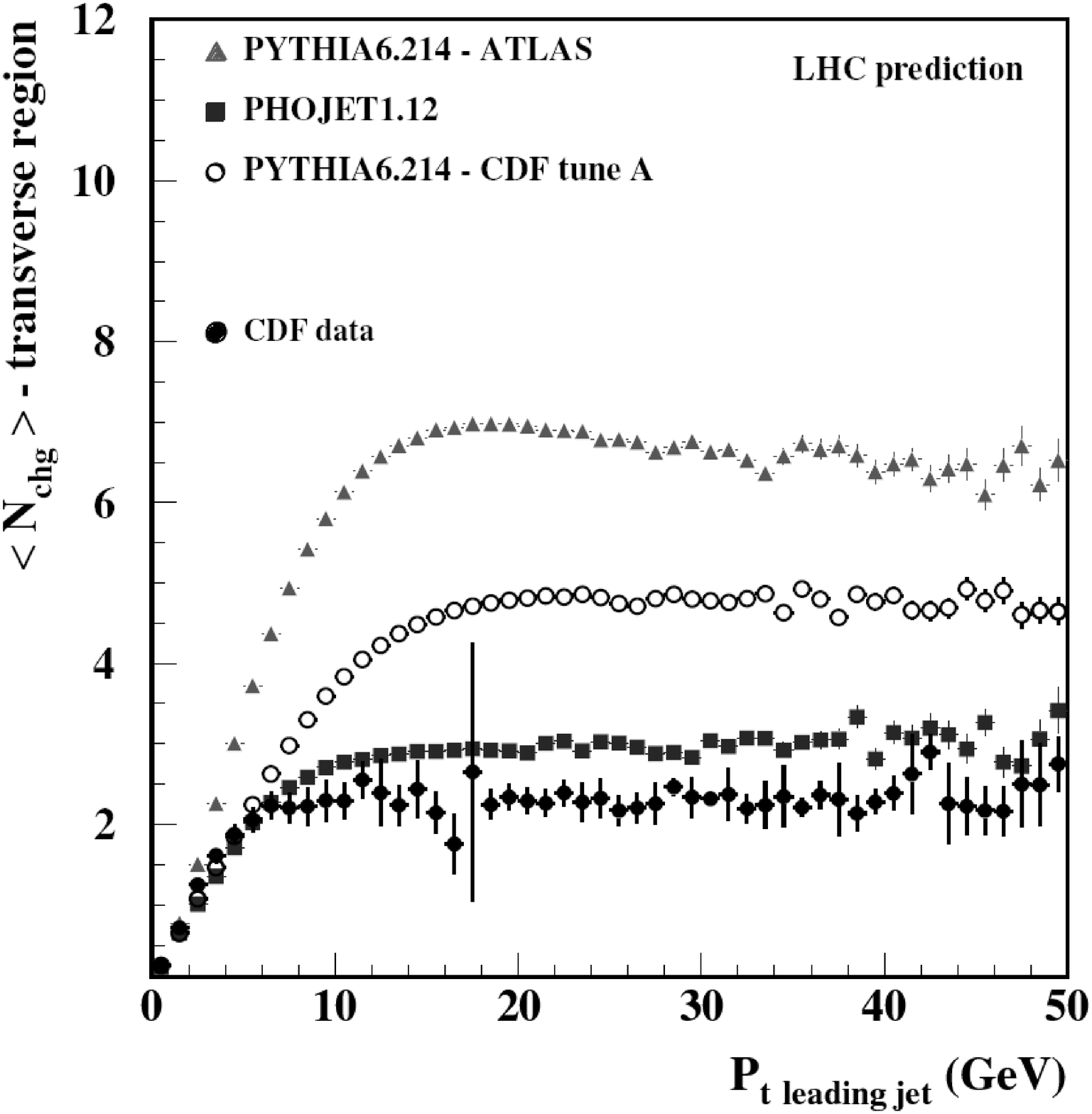}
  \caption{
    \label{fig:transv_lhc}
    Multiplicity in the transverse region for LHC runs with \hwpp\
    (left) and the same observable for several other generators (right),
    taken from Ref.~\cite{Alekhin:2005dx}. The different data sets for the left plot are (from
    bottom to top): Tevatron with MPI off, LHC with MPI off, Tevatron
    with MPI on and LHC with MPI on.}
\end{figure}

For calculating the LHC extrapolations we left the MPI parameters at
their default values, i.e.\ the fit to Tevatron CDF data. In
Ref.~\cite{Alekhin:2005dx} a comparison of different predictions for an
analysis modelled on the CDF one discussed earlier was presented. As a
benchmark observable the charged particle multiplicity in the transverse
region was used. We show this comparison in Fig.~\ref{fig:transv_lhc}
together with our simulation. All expectations reached a plateau in this
observable for $\pt^{\rm ljet} > 10$~GeV. Our prediction for this
observable also reached a roughly constant plateau within this
region. The height of this plateau can be used for comparison. In
Ref.~\cite{Alekhin:2005dx} PYTHIA 6.214 \cite{Sjostrand:2001yu} ATLAS
tune reached a height of $\sim 6.5$, PYTHIA 6.214 CDF Tune~A of $\sim 5$
and PHOJET 1.12 \cite{Engel:1994vs} of $\sim 3$.  Our model reaches a
height of $\sim 5$ and seems to be close to the PYTHIA 6.214 CDF tune,
although our model parameters were kept constant at their values
extracted from the fit to Tevatron data.

We have seen already in the previous section that our fit results in a
flat valley of parameter points, which all give a very good description
of the data. We will briefly estimate the spread of our LHC
expectations, using only parameter sets from this valley. The range of
predictions that we deduce will be the range that can be expected
assuming no energy dependence on our main parameters. Therefore, early
measurements could shed light on the potential energy dependence of the
input parameters by simply comparing first data to these predictions. We
extracted the average value of the two transverse observables for a
given parameter set in the region $20 \GeV < \pt^{\rm ljet} < 30
\GeV$. We did that for the best fit points at three different values for
$\ptmin$, namely 2 GeV, 3.4 GeV and 4.5 GeV, and found an uncertainty of
about 7 \% for the multiplicity and 10 \% for the sum of the transverse
momentum.

\begin{table}[h]
  \begin{center}
    \begin{tabular}{l|c|c}
      LHC predictions &$\langle N_{\rm chg}\rangle^{\rm transv}$ &
      $\langle \pt^{\rm sum}\rangle^{\rm transv} [\GeV]$\\ \hline TVT
      best fit& $5.1 \pm 0.3$ & $5.0 \pm 0.5$\\
    \end{tabular}
    \caption{LHC expectations for $\langle N_{\rm chg}\rangle$ and
      $\langle \pt^{\rm sum}\rangle$ in the transverse region. The
      uncertainties are obtained from varying $\ptmin$ within the range
      we considered. For $\mu^2$ we have taken the corresponding best
      fit (Tevatron) values.}\label{tab:LHCspread}
  \end{center}
\end{table}

\section*{Acknowledgements}
We would like to thank our collaborators on the \hwpp\ project for many
useful discussions. We wish to thank the organisers of the workshop for
a very pleasant atmosphere. This work was supported in part by the
European Union Marie Curie Research Training Network MCnet under
contract MRTN-CT-2006-035606 and the Helmholtz--Alliance ``Physics at
the Terascale''. MB was supported by the Landesgraduiertenf\"orderung
Baden-W\"urttemberg.
\bibliographystyle{JHEP} 
{\raggedright
\bibliography{lit}
}
\end{document}